\documentclass{aastex63}

\usepackage{graphicx}
\usepackage{amsmath}
\usepackage{amssymb}
\usepackage{romannum}
\submitjournal{PSJ}
\shorttitle{Transit Spectra of Titan at Ultraviolet Wavelengths}
\shortauthors{Tribbett et al.}
\begin{document}

\title{Titan in Transit: Ultraviolet Occultation Observations Reveal a Complex Atmospheric Structure}

\correspondingauthor{Patrick D. Tribbett}
\email{pdt43@nau.edu}

%
\author[0000-0001-9529-8353]{Patrick D. Tribbett}
\affiliation{Department of Astronomy and Planetary Science, Northern Arizona University, Box 6010, Flagstaff, AZ 86011, USA}
\affiliation{Habitability, Atmospheres, and Biosignatures Laboratory, Northern Arizona University, Flagstaff, AZ 86011, USA}

\author[0000-0002-3196-414X]{Tyler D. Robinson}
\affiliation{Department of Astronomy and Planetary Science, Northern Arizona University, Box 6010, Flagstaff, AZ 86011, USA}
\affiliation{Habitability, Atmospheres, and Biosignatures Laboratory, Northern Arizona University, Flagstaff, AZ 86011, USA}
\affiliation{NASA Astrobiology Institute’s Virtual Planetary Laboratory, University of Washington, Box 351580, Seattle, WA 98195, USA}

\author{Tommi T. Koskinen}
\affiliation{Lunar and Planetary Laboratory, University of Arizona, 1629 E. University Blvd., Tucson, AZ 85721, USA}
%

%
\begin{abstract}
    Transit spectroscopy is a key tool for exoplanet atmospheric characterization.  However, transit spectrum observations can be limited by aerosol extinction when gas opacities are weak.  The ultraviolet wavelength range contains a variety of strong molecular and atomic features, potentially enabling gas species detection even when atmospheric hazes are present.  To understand the interplay between aerosol extinction and ultraviolet molecular opacities, we investigate transmission through the atmosphere of Saturn's moon Titan during an occultation observed with the Ultraviolet Imaging Spectrometer (UVIS) aboard NASA's \textit{Cassini} orbiter.  We analyze the derived ultraviolet transit spectrum of Titan using atmospheric retrieval models that both include and exclude treatments for hazes.  Our retrieved atmospheric properties, namely the gas column densities, are consistent with previous studies analyzing UVIS occultation data.  Using the Bayesian Information Criterion, we demonstrate that haze parameterizations were unnecessary to fit the data despite apparent opacity due to multiple detached haze layers in the underlying occultation data.  Our work indicates that continued characterization of exoplanets in the ultraviolet wavelength regime can provide novel atmospheric constraints even if transit spectra are dominated by haze extinction at longer wavelengths.
\end{abstract}

%

%
%

%
\section{Introduction} \label{sec:intro}
%

Understanding the atmospheres of exoplanets provides essential insight into the formation, evolution, and potential habitability of these systems \citep{seager2010exoplanet,madhusudhan2019}.  Over the last two decades, transit spectroscopy \citep{seager2000theoretical,brown2001,hubbardetal2001} has emerged as the leading technique for characterizing exoplanet atmospheres.  Here, atmospheric opacity sources can lead to small variations in the wavelength-dependent dimming of a stellar host during an exoplanet transit event.  Despite the subtle nature of this effect, spectroscopic transit observations have yielded detections of atmospheric species in a diversity of exoplanet atmospheres \citep{charbonneau2002detection,tinettietal2007,swainetal2008,stevensonetal2010,lineetal2014,fraine2014water,singetal2016,bennekeetal2019,tsiarasetal2019}.

Of course, not all spectroscopic transit observations have revealed atmospheric features.  Especially for lower-mass exoplanets, observations have sometimes revealed flat, featureless transit spectra \citep{kreidberg2014clouds,knutson2014featureless,dewitetal2016}, at least to within measurement uncertainties.  Here, the presence of high altitude aerosols is often used to explain such flat transit spectra.  As the transit geometry implies that transit spectra probe long pathlengths along the limb of an exoplanet, even hazes or clouds with small vertical optical depths can appear opaque \citep{fortney2005effect}.

In theory, observing transit spectra in wavelength ranges where molecular or atomic opacities are relatively large will probe the upper reaches of exoplanet atmospheres, thereby potentially avoiding the obscuring effects of hazes and providing stronger detections of atmospheric species.  This, for example, leads to a key strength of NASA's upcoming {\it James Webb Space Telescope} \citep{gardneretal2006}, whose spectral coverage overlaps strong molecular rotation-vibration bands in the near- and mid-infrared with relevance to exoplanet transit spectra \citep{deming2009,beichmanetal2014,greeneetal2016,barstow2016,batalha&line2017}.  Another key wavelength regime with the potential for strong molecular and atomic opacities---the ultraviolet---was originally suggested as a range with likely high utility \citep{hubbardetal2001} and has recently been exploited to study the atmospheres of WASP-121~b \citep{sing2019hubble} and HAT-P-41~b \citep{wakeford2020uv}, both building on earlier efforts in the ultraviolet for HD~189733~b by \citet{sing2011hubble}.  In these works, \citet{sing2019hubble} observe strong Fe II and Mg II features for WASP-121~b, while both \citet{wakeford2020uv} and \citet{sing2011hubble} observe sloped transit spectra in the ultraviolet that are consistent with hazes for HAT-P-41~b and HD~189733~b, respectively.  In new modeling work, \citet{lothringer2020} detailed how strong opacities at ultraviolet wavelengths due to metals and metal-bearing species could help probe rainout chemistry in exoplanet atmospheres.

To further explore the interplay between aerosol extinction and absorption due to atmospheric gas species in ultraviolet transit observations, we turn to Titan.  For Titan, solar photons and wind, galactic cosmic rays, and magnetospheric charged particles drive pervasive atmospheric chemistry resulting in multitudes of higher order hydrocarbons and the carbon nitrogen aggregate tholins \citep{yung1984photochemistry, lavvas2008coupling, toublanc1995photochemical, podolak1979photochemical, carrasco2018evolution,  vuitton2009composition, lavvas2011energy}.  Additionally, Titan has a seasonally dependent detached haze layer located between 300 and 500~km altitude \citep[or at pressures lower than 10$^{-5}$~bar;][]{lavvas2009detached, west2011evolution, west2018seasonal} that is potentially analogous to the hazes responsible for some featureless exoplanet transit spectra.  However, it should be noted that due to drastically different thermal conditions, aerosol haze composition and particle sizes for hot exoplanets may differ from those seen in Titan's atmosphere \citep{lavvas2017aerosol, lavvas2019photochemical}.  For a recent review of Titan's atmosphere and climate, see \citet{horst2017titan}.

Here, we use Titan atmospheric stellar occultation observations from the Ultraviolet Imaging Spectrometer \citep{mcclintock1993optical, esposito2004cassini} aboard NASA's {\it Cassini} spacecraft to effectively study a hazy world in transit.  These occultation data are converted to exoplanet-like transit spectra following techniques developed by \citet{robinson2014titan} and \citet{dalbaetal2015}.  Critically, these occultation observations have already been used to derive key atmospheric properties for Titan, including number density profiles for various trace hydrocarbons \citep{koskinen2011mesosphere}, thereby helping to confirm aspects of our transit spectral analysis.  Moreover, it has been suggested that Titan's hydrocarbon rich atmosphere may be representative of a fairly common class of exoplanets, reinforcing the necessity of understanding the interplay of molecular opacities and haze extinction \citep{lunine2010titan}.

Below, we begin by describing our adopted occultation dataset and technique for converting this to an ultraviolet transit spectrum of Titan.  We then present the details of an atmospheric retrieval model designed to interpret our derived transit spectrum.  Following our retrieval analyses, we discuss the limited impact of Titan's detached haze layer on transit spectra.  Finally, we conclude by interpreting our results with respect to the current state of exoplanet observations.

%
\section{Methods} \label{sec:methods}
%

The following subsections describe our approach to data reduction, modeling, and analysis.  First, we briefly describe the underlying occultation dataset and how this was transformed into an effective transit spectrum for Titan.  Next, we present a simple forward model that we use to fit our ultraviolet transit spectrum of Titan.  Finally, we describe our Bayesian approach to atmospheric characterization using our simulated transit spectrum and forward model.

\subsection{Data Reduction and Transformation}
\begin{figure}
    \centering
    \includegraphics{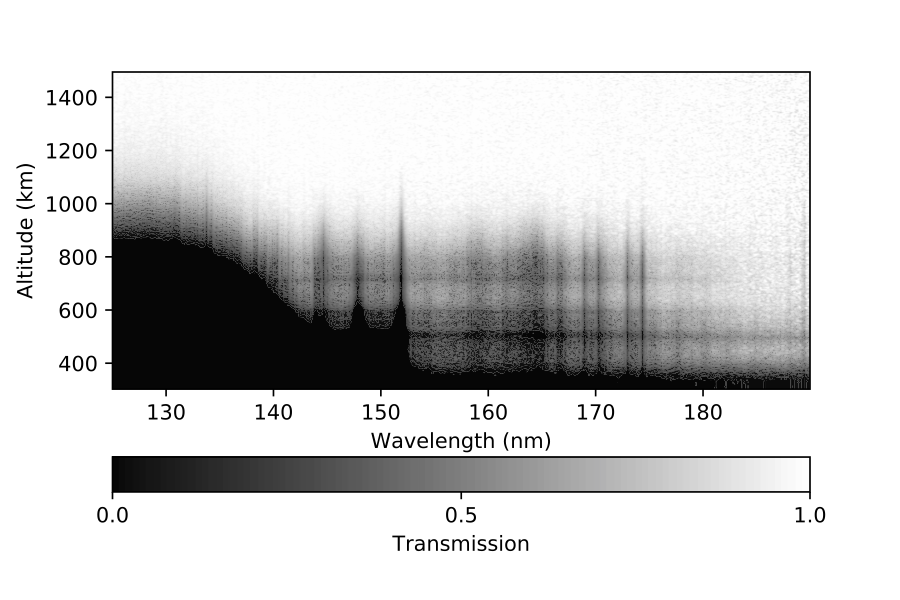}
    \caption{Altitude- and wavelength-dependent transmission data for Titan's atmosphere from {\it Cassini} flyby T41~\Romannum{1}.  An altitude of 0~km corresponds to Titan's average surface radius (2575~km), and darker colors indicate lower transmission.}
    \label{fig:trans_data}
\end{figure}

Occultation data were acquired with the far-ultraviolet channel of the Ultraviolet Imaging Spectrometer (UVIS) on board the NASA {\it Cassini} orbiter, as detailed in \citet{koskinen2011mesosphere}.  Spectra from the far-ultraviolet channel span 110--190~nm with a spectral resolution of 0.28~nm.  Further details regarding the optical specifications and design of the UVIS instrument are described in \citet{mcclintock1993optical}.  Data used in our analyses are from {\it Cassini} flyby T41~\Romannum{1} on 23 February 2008, where occultation observations probed 6$^{\circ}$S and 333$^{\circ}$W. Of the 12 independent occultation observations presented in \citet{koskinen2011mesosphere}, the T41~\Romannum{1} dataset is the best option for both high signal-to-noise-ratio data as well as a strong signature of two detached haze layers.  Transmissivity as a function of wavelength and altitude were calculated based on the ratio of the transmitted stellar spectrum along the instrument line of sight and the unocculted reference stellar spectrum.  This inherently requires that all spectral variations are atmospheric in nature. Quoted observational uncertainties are derived based on photon counting \citep{esposito2004cassini, koskinen2011mesosphere}.  The altitude- and wavelength-dependent transmission data from {\it Cassini} flyby T41~\Romannum{1} are shown in Figure~\ref{fig:trans_data}.  Note the two distinct vertically-isolated opacity sources in the atmosphere near 500~km and 700~km.  The lower region corresponds to the detached haze layer within Titan's atmosphere \citep{koskinen2011mesosphere}.  The upper region has also been attributed to higher order hydrocarbon haze, however the structure may be due to atmospheric propagation of gravity waves \citep{strobel2006gravitational, koskinen2011mesosphere}.

The transit depth spectrum corresponding to the altitude-dependent atmospheric transmission data was calculated following methods described in \citet{robinson2014titan}.  Given the wavelength-dependent transmissivity ($t_{\lambda,i}$) on a grid of impact parameters ($b_i$; taken as the radial distance of closest approach for a ray), the effective area of the Sun that would be blocked by an atmospheric annulus spanning levels $i$ to $i+1$ if Titan were observed in transit is given by,
\begin{equation}
    \Delta A_{\lambda,i} = \left(1 - \bar{t}_{\lambda,i} \right) \cdot \pi\left( b_{i+1}^{2} - b_{i}^{2} \right) \ ,
\end{equation}
where $\bar{t}_{\lambda,i}=\left(t_{\lambda,i+1}+t_{\lambda,i}\right)/2$. The wavelength-dependent transit depth is then computed by summing over annuli, with,
\begin{equation}
    \left( \frac{R_{{\rm p},\lambda}}{R_{\odot}}\right)^{\! 2} = \left( \frac{R_{\rm p}}{R_{\odot}}\right)^{\! 2} + \frac{1}{\pi R_{\odot}^2} \sum_{i}\Delta A_{\lambda,i} \ ,
\label{eqn:tdepth}
\end{equation}
where $R_{p}$ is the solid surface radius of the planet or, for worlds without a solid surface, a sufficiently deep reference radius that all transmissivity values have reached zero.  Corresponding transit depth uncertainties were calculated following standard Gaussian error propagation techniques \citep{taylor1997introduction}.  Equation~\ref{eqn:tdepth} can be solved for the wavelength-dependent planetary radius ($R^2_{{\rm p},\lambda}$), and subtracting the solid surface radius of Titan from this yields the so-called effective transit altitude (or height). Unlike in \citet{robinson2014titan}, losses due to refraction can be ignored here as the occultation observations probe much greater altitudes (i.e., very low pressures and number densities).  In all analyses, UVIS data between 185--190~nm are omitted due to higher order hydrocarbon  features, for which we do not have sufficient absorption cross section data \citep{koskinen2011mesosphere}.

\begin{figure}
    \centering
    \includegraphics[trim={0 6cm 0 6cm}, clip, width = 0.75\textwidth]{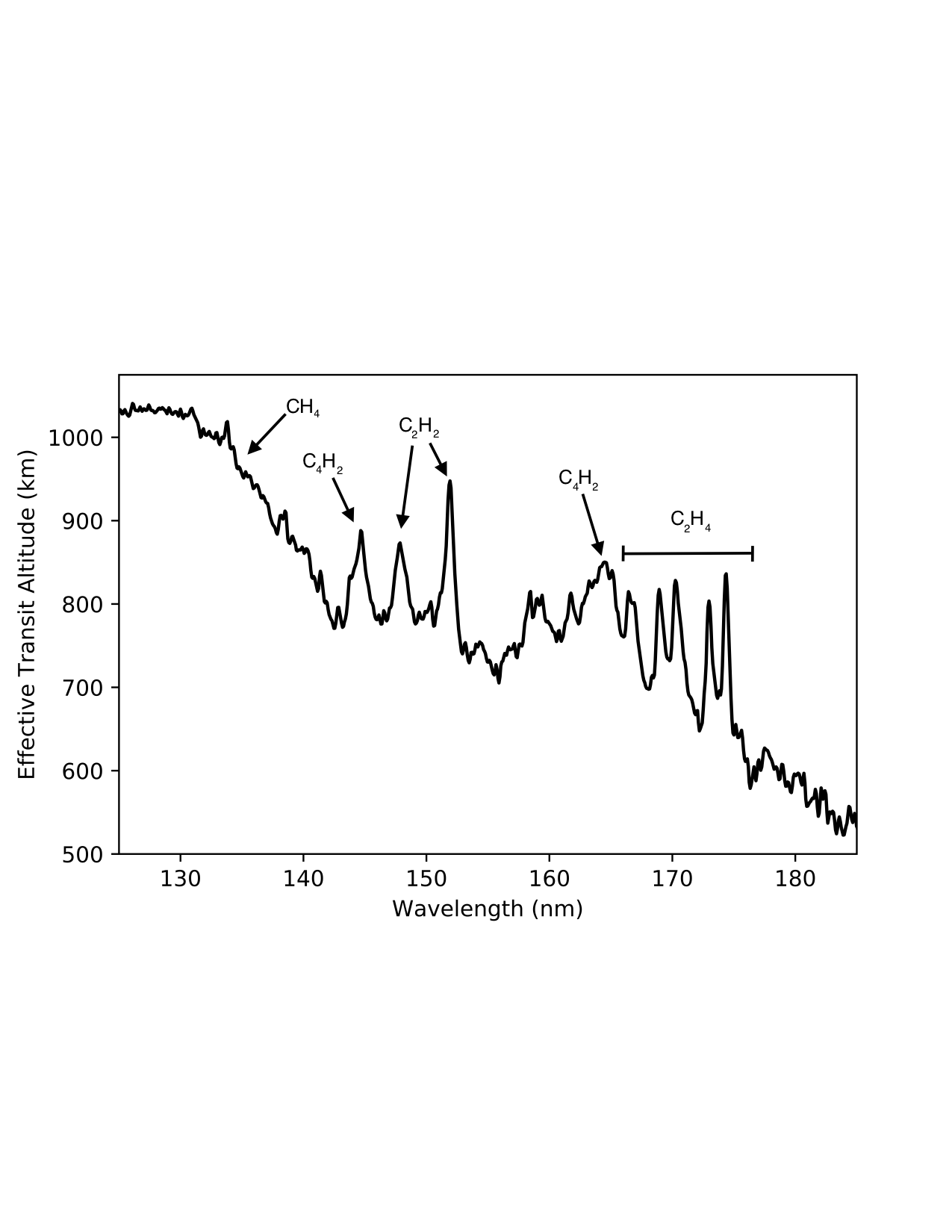}
    \caption{Effective transit altitude for Titan from {\it Cassini} flyby T41~\Romannum{1} occultation data.}
    \label{fig:zeff_spec}
\end{figure}

Figure~\ref{fig:zeff_spec} shows the transit spectrum that results from application of the previously described techniques, depicted here as effective transit altitude to help indicate where (vertically) in the atmosphere the transit spectrum probes.  Despite the detached haze layers present over this altitude regime, the modeled spectrum is rich with molecular features.  Several notable features include the broad methane absorption centered near 130~nm, acetylene and diacetylene features between 140~nm and 150~nm, and the ethylene features around 170~nm.

\subsection{Forward Model}
We seek to define a forward model that: (1) will enable retrievals of key atmospheric parameters when applied to our derived transit depth spectrum, and (2) is analogous in complexity to similar models for exoplanets.  Following \citet{benneke&seager2012} and \citet{robinson2014titan}, if the number density of an extincting species is distributed exponentially with scale height $H$ in an atmosphere, and if the extinction cross section for this species is pressure-independent, then the optical depth integrated along a slant path for an impact parameter $b$ and for a single species is given by
\begin{equation}
    \tau_{\lambda,j}(b) = 2 N_{0,j} \sigma_{\lambda,j} \frac{b}{H} K_{1}\left( \frac{b}{H} \right) e^{R_0/H} \ ,
\end{equation}
where a sub-script `$j$' indicates the species, $N_{0,j}$ is the vertical column number density above a sufficiently deep reference radius $R_0$, $\sigma_{\lambda,j}$ is the absorption cross section for the species, and $K_{n}$ is a modified Bessel function of the second kind.  Assuming all species are distributed with the same scale height, summing over all species yields the total slant optical depth, 
\begin{equation}
    \tau_{\lambda}(b) = 2\frac{b}{H} K_{1}\left( \frac{b}{H} \right) e^{R_0/H} \cdot \left( \sum_{j} N_{0,j} \sigma_{\lambda,j} \right) \  .
\end{equation}
Given the total slant optical depth, and assuming that extinction optical depth is equivalent to absorption optical depth \citep[see][]{robinsonetal2017}, the transit depth is obtained by integrating over impact parameters,
\begin{equation}
    \left(\dfrac{R_{{\rm p},\lambda}}{R_{\odot}}\right)^{\! 2} = \left(\dfrac{R_{0}}{R_{s}}\right)^{\! 2} + \frac{2}{R^2_{\odot}} \int_{R_{0}}^{\infty} \left[1 - e^{-\tau_{\lambda}\left( b \right)} \right] b db \ .
\end{equation}
Finally, for numerical implementation it is convenient to define a dimensionless parameter, $\beta = b/H$, so that,
\begin{equation}
    \tau_{\lambda}(\beta) = 2 \beta K_{1}\left( \beta \right) e^{R_0/H} \cdot \left( \sum_{j} N_{0,j} \sigma_{\lambda,j} \right) \  ,
\end{equation}
and,
\begin{equation}
    \left(\dfrac{R_{{\rm p},\lambda}}{R_{\odot}}\right)^{\! 2} = \left(\dfrac{R_{0}}{R_{s}}\right)^{\! 2} + 2\frac{H^2}{R^2_{\odot}} \int_{R_{0}/H}^{\infty} \left[1 - e^{-\tau_{\lambda}\left( \beta \right)} \right] \beta d\beta \ ,
\end{equation}
where a grid of $\beta$ values can be straightforwardly designed to ensure vertical resolution that is (at least) finer than $H$.

\begin{figure}
    \centering
    \includegraphics{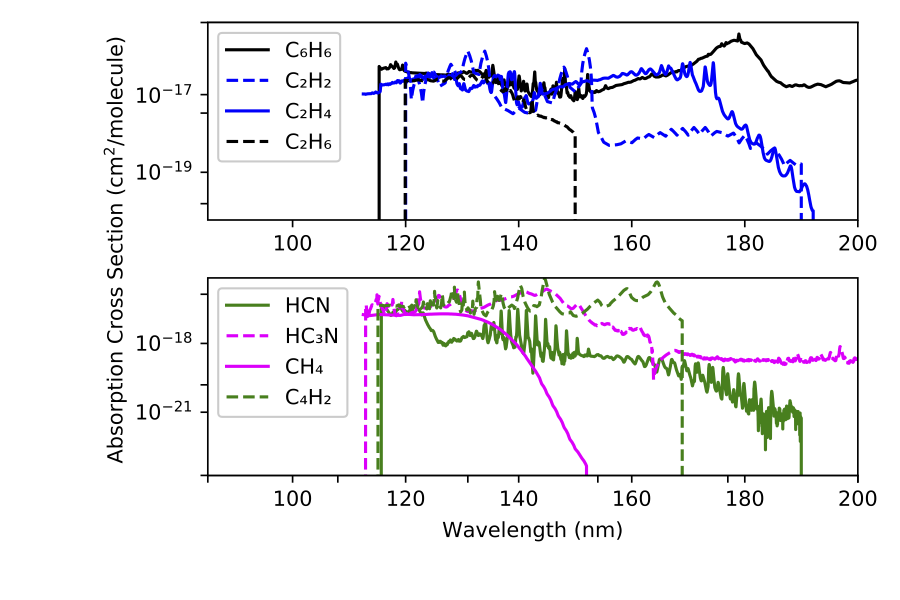}
    \caption{Absorption cross sections for the hydrocarbons and nitriles of interest in this study.}
    \label{fig:opacities}
\end{figure}

Following \citet{koskinen2011mesosphere}, our gaseous opacity sources emphasize hydrocarbon and nitrile species of substantial number densities in Titan's atmosphere and that have non-negligible absorption cross sections at deep ultraviolet wavelengths.  Species used in this study are shown in Table~\ref{tab:species} along with wavelength coverage, measurement temperature, and a reference paper for our adopted opacity data.  Opacity data were selected on the basis of wavelength range and temperature relevance.  Absorption cross sections for the eight species used in this study are shown in Figure~\ref{fig:opacities}.  To treat opaque atmospheric haze layers within our forward model, we defined an altitude below which the optical depth is inflated to simulate complete extinction.

Finally, as we seek, in part, to understand if Titan's detached haze layer significantly impacts the UVIS-derived transit spectrum, we explore two simple haze treatments in our forward model.  Our two-parameter haze model specifies the lower boundary of a haze layer ($z_{\rm h}$, measured above $R_0$) as well as a grey haze vertical optical depth $\tau_{\rm h}$, assumes that the optical depth is distributed uniformly over one scale height, and computes slant optical depths following the path distribution approach presented in \citet{robinson2017}.  A one-parameter haze model simply assumes the atmosphere is opaque for impact parameters that probe below $z_{\rm h}$ \citep[see, e.g.,][]{betremieux&swain2017,kempton2017}, which serves to block all light that would probe atmospheric layers deeper than the haze altitude and, as we show later, is instructive for interpreting results from our two-parameter haze treatment.

\begin{table}
\centering
\def\arraystretch{1.5}
 \begin{tabular}{|c c c c|} 
 \hline
 Species & Wavelength (nm) & Temperature (K) &  Reference\\
 \hline
 \hline
 Benzene (C$_{6}$H$_{6}$) & 115--205 & 298 &  \citet{capalbo2016new}\\
 Acetylene (C$_{2}$H$_{2}$) & 120--190 & 150 & \citet{wu2001measurements}\\
 Ethylene (C$_{2}$H$_{4}$) & 105--115 & 298 & \citet{lu2004analysis}\\
 & 115--192 & 140 & \citet{wu2004temperature}\\
 Ethane (C$_{2}$H$_{6}$) & 120--150 & 150 & \citet{chen2004temperature}\\
 Methane (CH$_{4}$) & 102--119 & 298 & \citet{ditchburn1955absorption}\\
 & 120--142 & 150 & \citet{chen2004temperature}\\
 & 143--152 & 298 & \citet{lee2001enhancement}\\
 Hydrogen Cyanide (HCN) & 115--190 & 255 & \citet{koskinen2011mesosphere} \\
 Cyanoacetylene (HC$_{3}$N) & 112--230 & 298 & \citet{koskinen2011mesosphere}\\
 Diacetylene (C$_{4}$H$_{2}$) & 115--168 & 173 & \citet{ferradaz2009temperature}\\
 [0.5ex] 
 \hline
 \end{tabular}
 \caption{Absorbing species included in our forward model along with wavelength range, measurement temperature, and a reference for the opacity data.  Opacities were derived from a compilation by \citet{essd-5-365-2013}.}
 \label{tab:species}
\end{table}

\subsection{Markov Chain Monte Carlo Implementation}
We retrieve atmospheric parameters by fitting our forward model to our UVIS-derived transit spectrum using a standard Mark chain Monte Carlo (MCMC) approach to Bayseian inference.  Specifically, we adopt the widely-used MCMC tool {\tt emcee}, developed by \citet{foreman2013emcee}.  Bayesian statistics describe the posterior conditional probability distribution of a set of parameters using the prior probabilities of those same parameters and their likelihood probabilities given the data.  The posterior distribution enables inference of physical parameters as well as uncertainties on these parameters.

Briefly, Bayes' Theorem states,
\begin{equation}
    P(\theta|y) = \dfrac{P(y|\theta) P(\theta)}{P(y)} \ ,
\end{equation}
or
\begin{equation}
    P(\theta|y) \propto P(y|\theta)P(\theta) \ ,
\end{equation}
where $y$ are the observed data, $\theta$ is our set of atmospheric parameters (i.e., our atmospheric state parameters), $P(\theta)$ is the prior probability for these parameters, and $P(y|\theta)$ is the so-called likelihood function.  The value of the likelihood function depends on the data, its associated errors, and the atmospheric state parameters.  When evaluating the likelihood function, parameter values that result in poorly fit data are penalized.  In our approach, the log of the likelihood function (often called the log-likelihood) is given by,
\begin{equation}
    \ln P(y|\theta) = - \dfrac{1}{2}\sum_{k} \dfrac{\left[ y_{k} - f(\theta) \right]^{2}}{\sigma_{k}^{2}} + \ln 2\pi\sigma_{k}^{2} \ ,
\end{equation}
where $y_{k}$ is the kth spectral data point, $\sigma_{k}$ is the uncertainty for this data point, and $f(\theta)$ is the transit depth (forward) model.  We adopt uninformed prior probability distributions that simply set physical limits on the values that each parameter can take.  Model free parameters, their descriptions, and corresponding prior probability limits for our analyses are shown in Table~\ref{tab:params}.  Our prior probabilities are largely based on the physical limits of the properties that the parameters represent (e.g., all lengths cannot be smaller than zero).  Column number densities are retrieved in log space to ensure a more full exploration of parameter space over multiple orders of magnitude.

\begin{table}
\centering
\def\arraystretch{1.5}
 \begin{tabular}{|c c c c|} 
 \hline
 Model Parameter &  Description & Units & Prior\\
 \hline
 \hline
 $H$ & Scale Height & km & $0 < H$ \\
 $R_{0}$ & Reference Radius & km & $0 < R_{0}$ \\ 
 $\log N_{j} $ & Column Number Density for Species j & $\log {\rm cm}^{-2}$ & $12 < \log N_{j} < 24$\\
 $z_{\rm h}$ & Haze Layer Altitude above $R_{0}$ & km &  $0 < z_{\rm h}$ \\
 $\tau_{\rm h}$ & Haze Layer Optical Depth & - & $0 < \tau_{\rm h} < 10$\\
 [0.5ex] 
 \hline
 \end{tabular}
 \caption{Model parameters, their units, and our adopted priors.  Column number densities are measured vertically above $R_{0}$. }
 \label{tab:params}
\end{table}

In later sections, We use the Bayesian Information Criterion (BIC) to determine if the addition of a haze layer is warranted in our forward model.  The BIC has become increasingly common in the astronomy community by providing a robust means for model comparisons \citep[e.g.,][]{littenberg2009bayesian,feng2016impact,sharma2017markov}, and is defined by,
\begin{equation}
    {\rm BIC} = -2 \ln P\left(y\vert\theta \right)_{\rm max} + \nu \ln N
\label{eqn:bic}
\end{equation}
where $\nu$ is the number of model free parameters, $N$ is the number of spectral data points, and $P(y\vert\theta)_{\rm max}$ is the maximized log likelihood.  The change in the BIC, $\Delta{\rm BIC}$, between two models is then evaluated to support or reject adding additional parameters (e.g., $z_{\rm h}$ and $\tau_{\rm h}$ in our case) to improve the model fits.  The model with a lower BIC value is preferred over another.  Following \citet{kass1995bayes}, $\Delta$BIC values between 0 and 2 suggest that the evidence against the higher BIC model is ``not worth more than a bare mention."  Values between 2 and 6 suggest positive evidence against the larger BIC model.  Values between 6 and 10 suggest strong evidence against the larger BIC model.  Values greater than 10 suggest very strong evidence against the larger BIC model.  A more detailed description of the BIC and how to compare models can be found in \citet{kass1995bayes}.

%
\section{Results} \label{sec:results}
%

Clearsky (i.e., haze-free) and hazy models were applied within our retrieval framework to our derived transit spectrum of Titan.  As discussed immediately below, retrievals were initially performed assuming only photon counting errors in the observations.  Later retrievals --- more analogous to an exoplanet case --- were performed with inflated error bars.

\subsection{Error Scaling}
\begin{figure}
    \centering
    \includegraphics[width = 0.75\textwidth]{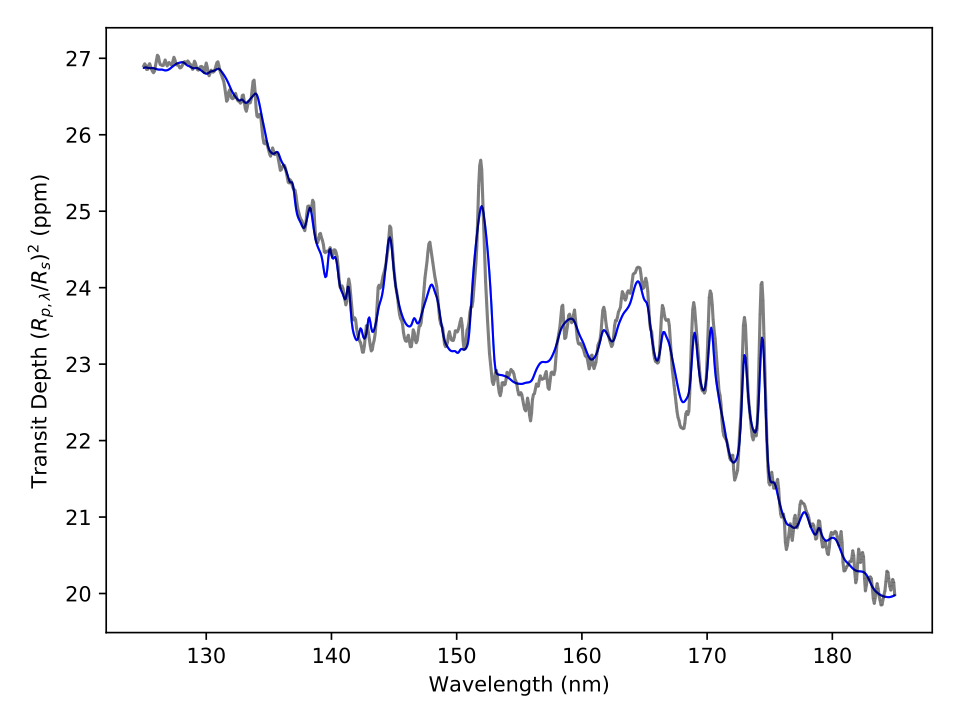}
    \caption{Titan transit depth spectrum is shown in black.  Propagated error bars are included; however, they are small enough such that they are not visible on this scale.  Modeled transit spectra from our clearsky retrieval analysis are shown as 1-$\sigma$ (68.2\%) and 2-$\sigma$ (95.5\%) spreads, dark blue and light blue respectively.}
    \label{fig:spectra_noerror}
\end{figure}

We initially applied our clearsky and two-parameter haze retrieval models to the UVIS-derived transit spectrum of Titan with only propagated photon counting errors.  Figure~\ref{fig:spectra_noerror} shows fitted spectra with uncertainties for the clearsky case.  While the modeled spectra appear qualitatively reasonable, the reduced chi-squared for the clearsky best-fit model was $2.7 \times 10^{5}$.  Similarly, the reduced chi-squared for the two parameter haze best-fit model was $2.7 \times 10^{5}$.  Here, quantitatively poor fits likely stem from a combination of incomplete opacity data and overly-simplified model assumptions.  For the former, and as can be seen in Table~\ref{tab:species}, the laboratory opacity data are often measured in a temperature regime that is much warmer than Titan's upper atmosphere (which has characteristic temperatures of 150--200~K).  Regarding model assumptions, previous analyses of the UVIS occultation observations reveal that the absorbing gaseous species do not have number density profiles that follow simple exponential decreases with a uniform scale height \citep{koskinen2011mesosphere}, as is assumed in our model.

\begin{figure}
    \centering
    \includegraphics{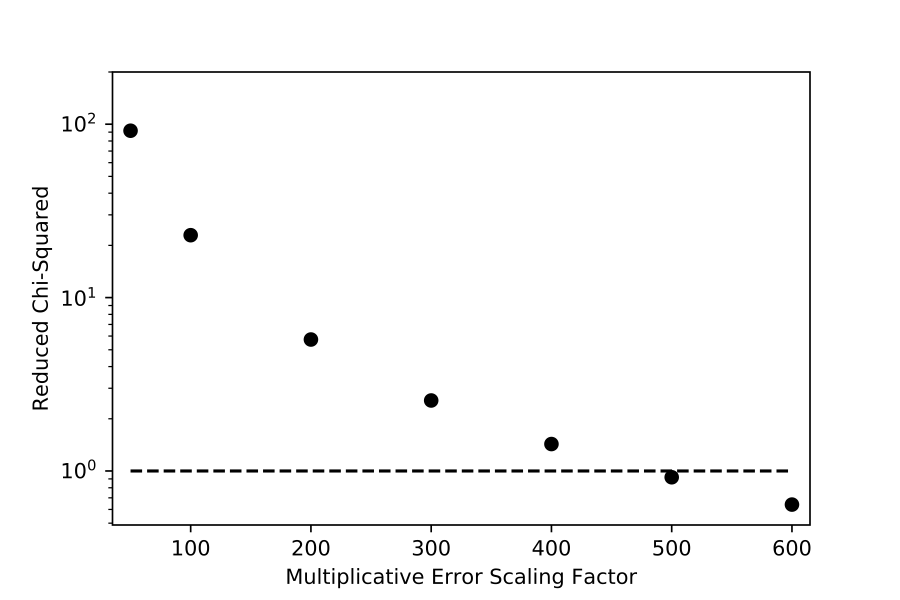}
    \caption{Reduced chi-squared values for the best-fit model from our retrieval framework when increasing the multiplicative error scaling factor on the UVIS-derived transit spectrum.  The dotted line represents a reduced chi-squared of unity.}
    \label{fig:scaling}
\end{figure}

To produce better fits (as indicated by the reduced chi-squared) and to better mimic retrieval analyses applied to exoplanet data, we opted to inflate the error bars on our transit spectrum via a uniform multiplicative scaling factor \citep[as is a relatively common practice; ][]{tremaine2002slope, hogg2010data, foreman2013emcee, line2015uniform}.  Figure~\ref{fig:scaling} demonstrates how increasing the data uncertainty affects the resulting reduced chi-squared for a best-fit clearsky model.  Based on this analysis, we chose a multiplicative scaling factor of 500 to produce reduced chi-squared values reliably close to unity without losing spectral information and potentially overfitting the data.

\subsection{Clearsky Retrievals}
\begin{figure}
    \centering
    \includegraphics[width = \textwidth]{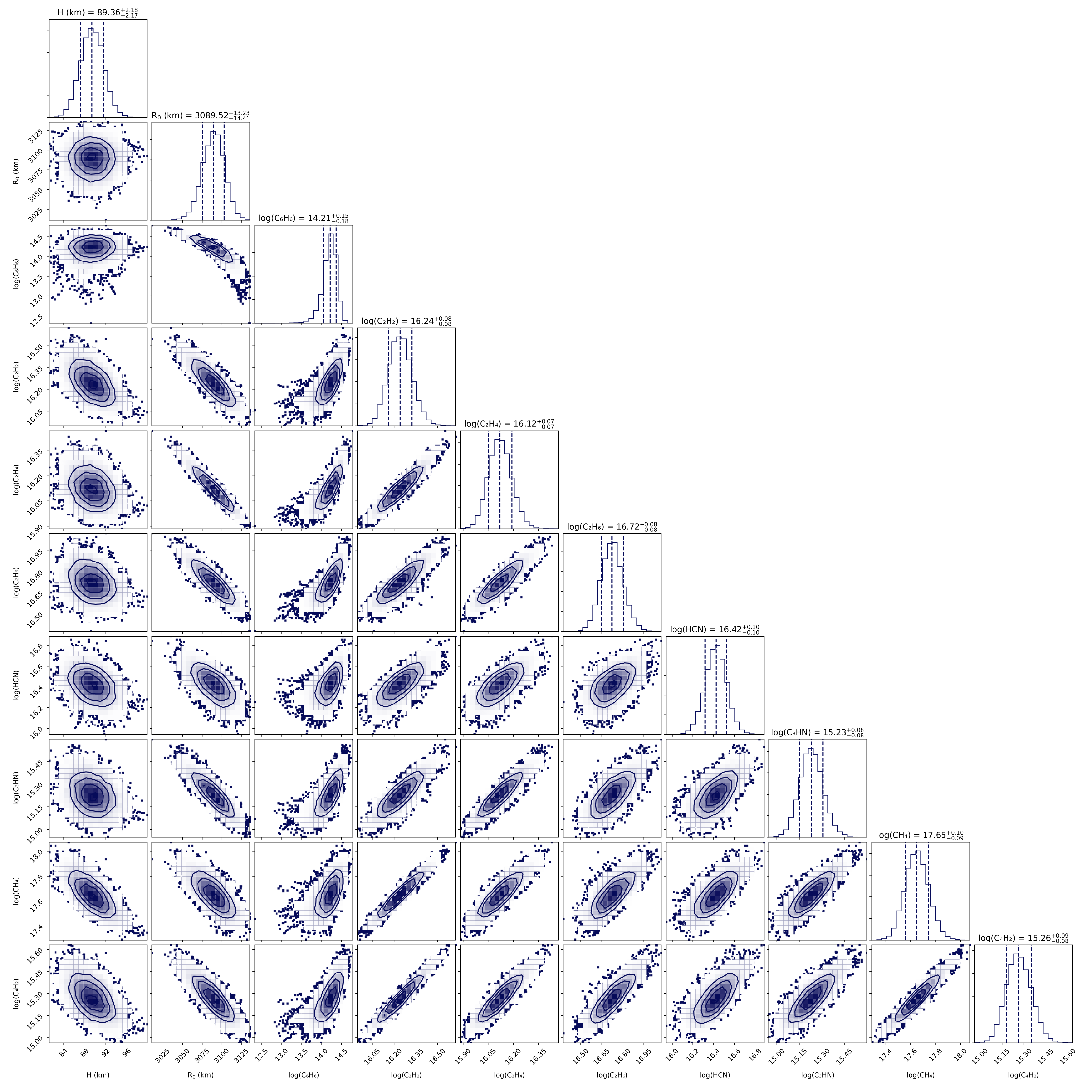}
    \caption{Marginalized posterior distributions for a clearsky retrieval analysis applied to the UVIS-derived transit data for Titan in Figure~\ref{fig:zeff_spec}. Sub-plots along the diagonal show posterior distributions where all but a single parameter have been marginalized over, and distribution values at the 16th, 50th, and 84th percentile are indicated above each sub-plot.  Off-diagonal sub-plots show posterior distributions where all but two parameters have been marginalized over.}
    \label{fig:corners_clear}
\end{figure}

\begin{figure}
    \centering
    \includegraphics[width = 0.75\textwidth]{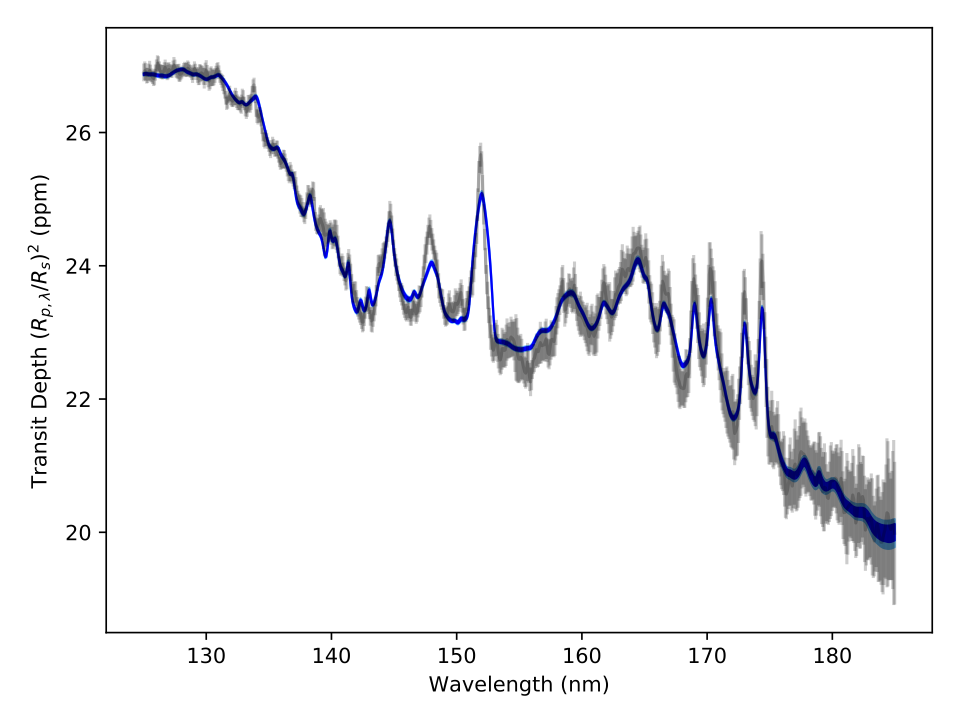}
    \caption{Titan transit depth spectrum (black) with $500\times$ inflated error bars (grey).  Modeled transit spectra from our clearsky retrieval analysis are shown as 1-$\sigma$ (68.2\%) and 2-$\sigma$ (95.5\%) spreads, dark blue and light blue respectively.}
    \label{fig:spectra_clear}
\end{figure}

Retrieved atmospheric parameters for the application of our clearsky model to the UVIS Titan transit spectrum with inflated uncertainties are shown in Figure~\ref{fig:corners_clear}.  This so-called ``corners'' plot depicts, along the diagonal, the posterior distributions for all retrieved parameters marginalized over all other parameters.  Off-axis plots show two-dimensional posterior distributions where all but two retrieved parameters have been marginalized over.  For the one-dimensional marginal distributions, values at the 16th, 50th, and 84th percentile are indicated above each sub-plot (i.e., the distribution mean and $\pm$1-$\sigma$), and the reduced chi-squared for the best-fit model was 0.917.  All one-dimensional marginal distributions are roughly Gaussian in shape.  The reference radius ($R_0$) is anti-correlated with all gas column number densities over a narrow range of parameter space as roughly fixed number densities aloft can be maintained when the reference radius is increased but the column number densities down to this radius are decreased.  These anti-correlations with $R_0$ then lead to correlations between all gas column number densities over narrow ranges of parameter space.  Finally, Figure~\ref{fig:spectra_clear} shows the 1-$\sigma$ (68.2\%) and 2-$\sigma$ (95.5\%) spread in model spectra derived from our retrieval.

\subsection{Hazy Retrievals}
\begin{figure}
    \centering
    \includegraphics[width = \textwidth]{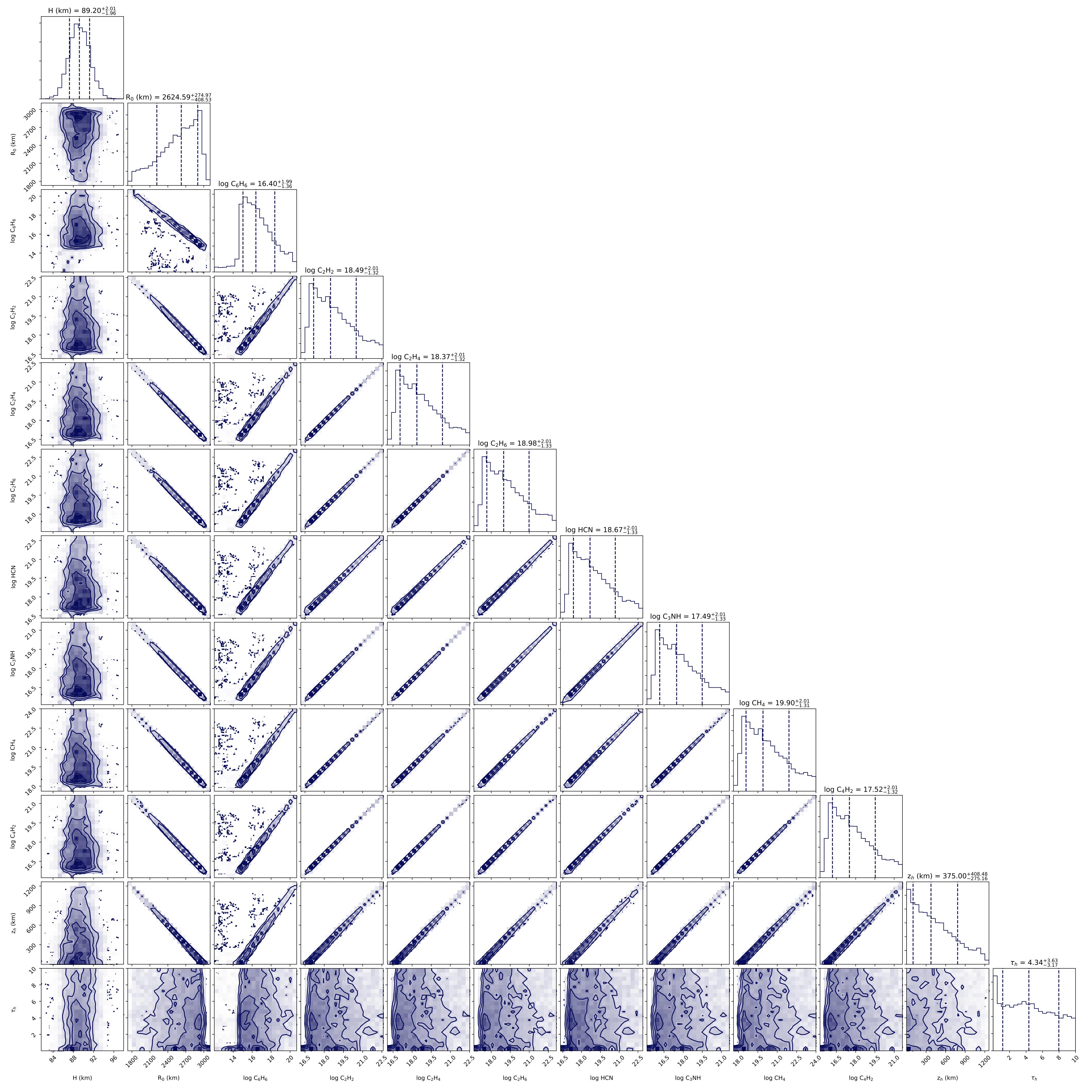}
    \caption{Same as Figure~\ref{fig:corners_clear} except with the addition of a two-parameter haze treatment.}
    \label{fig:corners_hazy2}
\end{figure}

\begin{figure}
    \centering
    \includegraphics[width = 0.75\textwidth]{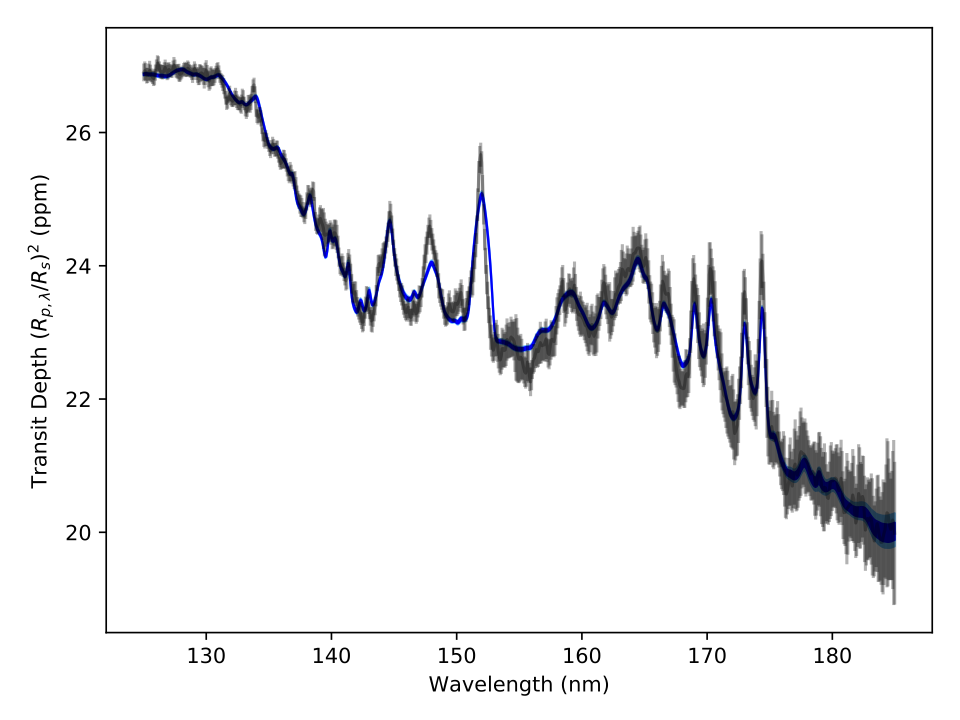}
    \caption{Same as Figure~\ref{fig:spectra_clear} except for retrieved spectra that include a two-parameter haze treatment.}
    \label{fig:spectra_hazy2}
\end{figure}

Marginalized posterior distributions for our two-parameter haze model applied to the UVIS-derived transit spectrum are shown in Figure~\ref{fig:corners_hazy2}.  The resultant spectra are shown in Figure~\ref{fig:spectra_hazy2}.  As can be seen in the spectra, fits produced by the two-parameter haze model show little to negligible difference in quality as compared to the clearsky model.  More quantitatively, the reduced chi-squared for the best-fit two-parameter haze model is 0.919, indicating that the models are nearly identical in their ability to fit the data.

Correlations between gas column number densities and the reference radius parameter in the two-parameter hazy model analysis are similar to those seen in the clearsky analysis.  However, and as compared to the clearsky analysis, the two-parameter hazy one-dimensional posterior distributions are generally wider and non-Gaussian (excepting the scale height distribution).  The detached haze vertical optical depth ($\tau_{\rm h}$) is poorly constrained, but is generally found to be optically thick in the horizontal (i.e., slant) direction.  A tail in the marginal distribution for the reference radius ($R_0$) to smaller radii correlates with $z_{\rm h}$ to maintain the floor of the transit spectrum near 3,100~km, or about 500~km above the solid body radius of Titan, which corresponds to the deepest altitudes probed in our transit spectrum.  This, in turn, leads to tails in the marginal distributions for the gas vertical column abundances towards larger values.  Lastly, it should be noted that the primary benzene feature ($\sim$ 180 nm) probes near 500 km in the transit spectrum.  For this reason, the distribution is likely impacted by the haze parameterization, which sets the floor of the transit spectrum near 500 km.  

\begin{figure}
    \centering
    \includegraphics[width = \textwidth]{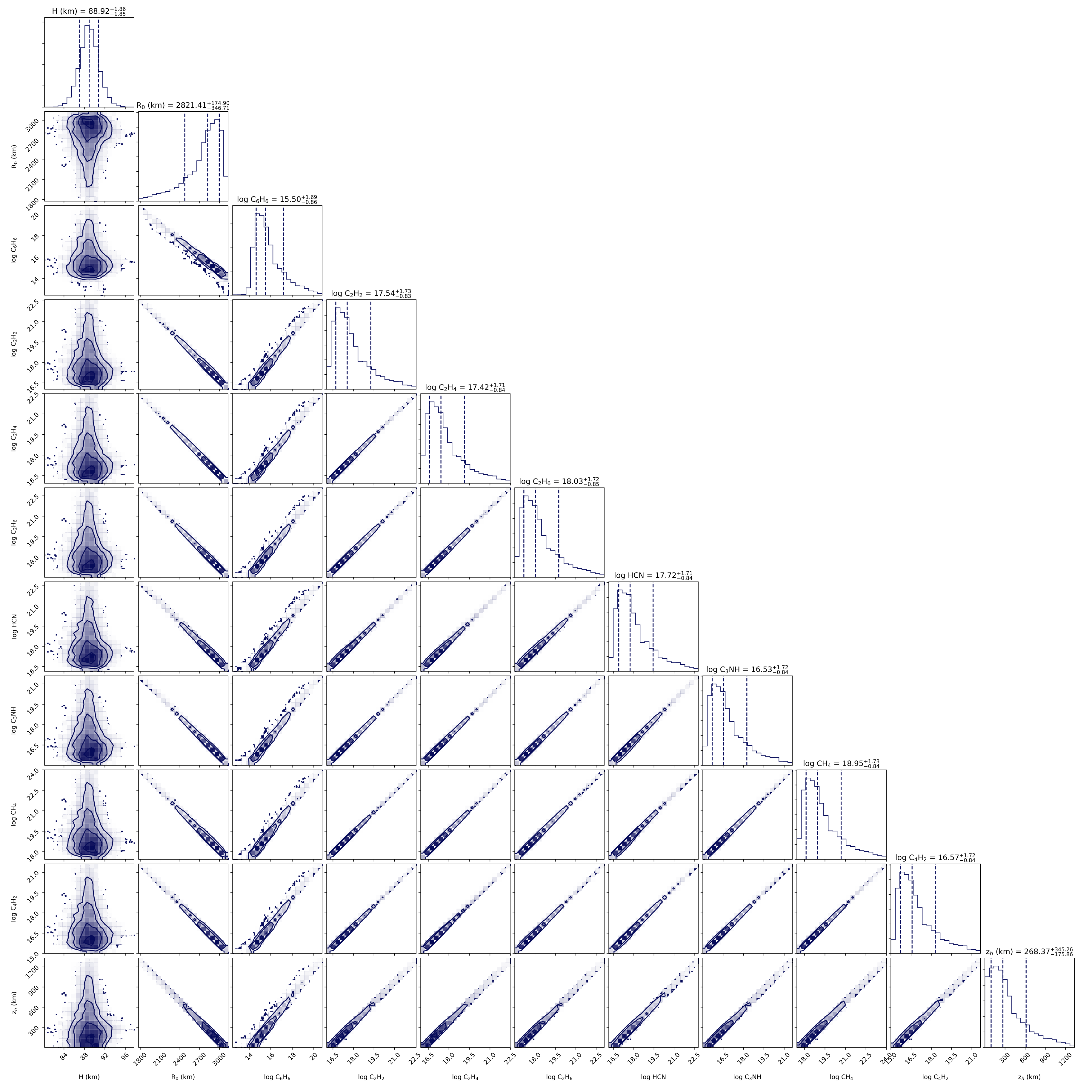}
    \caption{Same as Figure~\ref{fig:corners_clear} except with the addition of a single-parameter haze treatment.}
    \label{fig:corners_hazy1}
\end{figure}

\begin{figure}
    \centering
    \includegraphics[width = 0.75\textwidth]{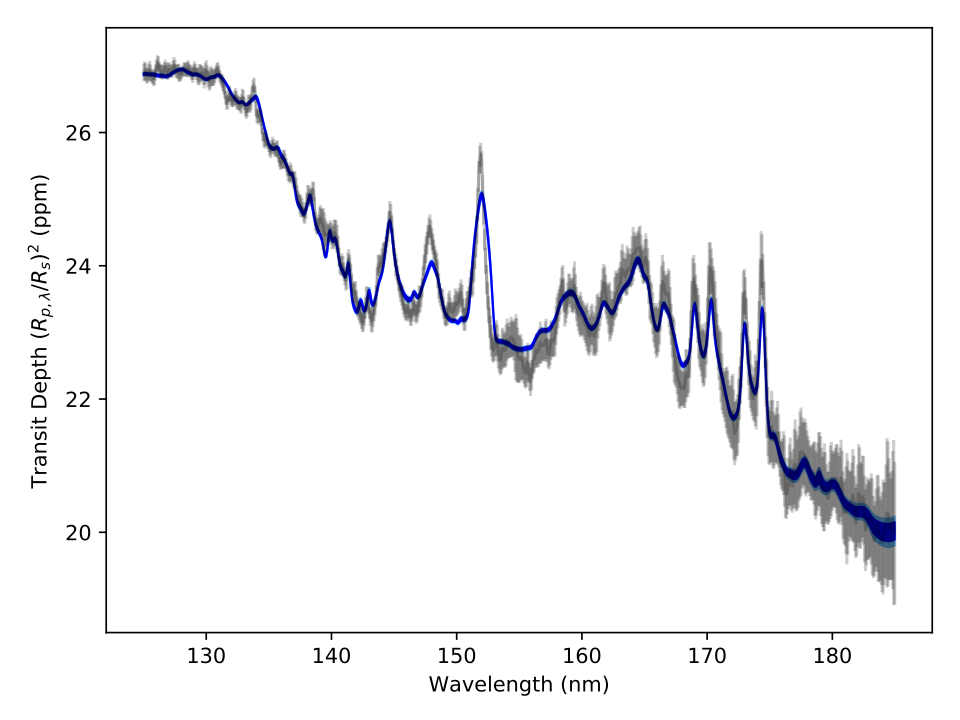}
    \caption{Same as Figure~\ref{fig:spectra_clear} except for retrieved spectra that include a single-parameter haze treatment.}
    \label{fig:spectra_hazy1}
\end{figure}

Results from our one-parameter haze model can be used to further understand the correlations and wide distributions seen in our two-parameter haze treatment.  Marginalized posterior distributions from the application of our one-parameter model are shown in Figure~\ref{fig:corners_hazy1} and the resulting spread in model spectra are shown in Figure~\ref{fig:spectra_hazy1}.  Here, the reference radius and haze altitude combine to set the floor of the transit spectrum near 500~km above Titan's solid body radius.  With the haze layer fixing the floor of the transit spectrum, $R_0$ can take on a wide range of values and, correspondingly, the column number densities also vary over a wide range of values.  These behaviors are also seen in the two-parameter haze model results and are analogous to a degeneracy between the reference radius, pressure, and gas abundances discussed for retrievals applied to transiting exoplanets in \citet{heng&kitzmann2017}.

%
\section{Discussion} \label{sec:discuss}
%

To compare with previous studies, we convert our vertical column number densities (measured above $R_0$) to slant path column number densities.  These two column number densities are related through,
\begin{equation}
\label{equ:ColConv}
    N_{s}(z) = \frac{2\left(R_{\rm p} + z \right)}{H} \cdot K_{1}\left( \frac{R_{\rm p} + z}{H}\right) \cdot e^{R_{0}/H} \cdot N_{0}
\end{equation}
where $N_{\rm s}$ is the slant path column number density of some species at a tangent altitude of $z$ above the solid body radius.  Using Equation~\ref{equ:ColConv}, we randomly re-sampled our MCMC-derived distributions for each species to derive slant column number density distributions at 700~km altitude \citep[selected to best compare to results in][]{koskinen2011mesosphere}.  With these derived distributions, we directly compare our slant column number densities to previous {\it Cassini}/UVIS retrievals.

Critically, our retrieved column number densities generally agree with previous studies that used {\it Cassini}/UVIS observations, as shown in Table~\ref{tab:comparison}. Thus, with sufficient signal-to-noise, transit observations can constrain atmospheric properties with uncertainties comparable to orbital measurements. Despite general agreement between our retrievals, \citet{koskinen2011mesosphere}, and \citet{shemansky2005cassini}, column number densities for ethane (C$_{2}$H$_{6}$) and benzene (C$_{6}$H$_{6}$) show more substantial deviation and merit some discussion.  Ethane is a known photochemical product within Titan's atmosphere \citep{lavvas2008coupling}.  Unfortunately, there are no apparent distinct ethane features within Titan's transit spectrum.  Rather, it provides relatively grey opacity between 120--135~nm, with a slight slope longward.  Using photochemical calculations and mixing ratio constraints \citet{koskinen2011mesosphere} estimate an upper limit for the column number density of ethane.  Our retrieved column number density, which is obtained without photochemical modeling, is in agreement to within an order of magnitude of previous UVIS results.  Additionally, retrieved column number densities for benzene show some variation.  These discrepancies likely stem from simplifying assumptions in our model --- namely that atmospheric species are distributed exponentially with scale height.  \citet{koskinen2011mesosphere} demonstrate that column densities for benzene have additional structure with altitude, especially near 700 km.  Generally, all retrieved column number density values across studies are fairly consistent.  However, it is important to also note the limitations for these comparisons, particularly in spectral regions where the extent to which the transit probes is limited.  For example, in the far UV ($\sim 130$nm) the transit spectrum probes much greater altitudes than the comparison altitude of 700 km.  As methane is responsible for the majority of the opacity in this region, our methane constraint is primarily applicable at altitudes markedly larger than 700 km and our assumption of a constant scale height for number density distributions leads to a discrepancy at 700 km.

Additionally, it is notable that the retrieved vertical column number densities for the one- and two-parameter haze models (shown in Figure~\ref{fig:corners_hazy2}) are generally biased larger than the clearsky values and are poorly constrained (Figure~\ref{fig:corners_clear}).  Though the extent to which these distributions are poorly constrained did not alter the slant column densities,  these distributions reinforce the importance of physically appropriate haze parameterizations and prior constraints on planetary radii.

To best inter-compare the utility of our clearsky and hazy models, we calculated the BIC for each model using Equation~\ref{eqn:bic}.  The $\Delta$BIC value was 7 when comparing the clearsky and one-parameter haze model, with the higher individual BIC corresponding to the hazy model.  This evidence suggests that there is no reason to include an additional free parameter to fit for an opaque haze layer.  For the clearsky and two-parameter haze model comparison, the $\Delta$BIC value was 13.  This indicates strong evidence that supports excluding the two-parameter detached haze layer treatment.  These findings remain true for data even without inflated error bars, indicating that error inflation did not result in substantive losses in spectral information.

It is tempting to interpret results from our two-parameter haze treatment in Figure~\ref{fig:corners_hazy2} as showing a slight preference for models with a detached haze layer near 500~km above Titan's solid body radius with a slant optical depth near unity (i.e., $\tau_{\rm h}$ near 0.1).  This would correspond to the haze layer near 500~km seen in the altitude-resolved transmission data (Figure~\ref{fig:trans_data}).  Future work might find stronger evidence for the detection of this detached haze layer (and, potentially, the haze layer near 700~km) in our transit spectrum by including wavelength-dependent (i.e., non-grey) opacity data for organic hazes and/or exploring haze models that do not assume a uniform distribution of aerosol opacity across a scale height.  Regarding the latter point, the transmission data in Figure~\ref{fig:trans_data} show a sharper increase in aerosol number density near the base of the haze layer near 500~km \citep{koskinen2011mesosphere}.  Along these lines, future work might also investigate how atmospheric constraints are degraded as observational uncertainties are artificially inflated to increasingly larger levels than are investigated here.

While our results demonstrate the rich atmospheric information that can be provided by observations of exoplanet transits at ultraviolet wavelengths, the acquisition of such data comes with additional complications.  Perhaps most fundamentally, the overall faintness of Sun-like and cool stars at ultraviolet wavelengths implies that reaching even 10--100~ppm accuracy on transit spectrum observations is difficult.  Specifically, shortward of 140 nm, Sun-like stellar spectra exclusively consist of atomic emission lines and minimal continuum severely limiting the SNR \citep{young2018structure}.  Beyond this, the so-called ``transit light source effect'' will be most pronounced at short wavelengths \citep[see, e.g.,][]{rackhametal2018}.  Here, occulted heterogeneities on the stellar photosphere --- namely spots and faculae --- can introduce systematic biases in transit spectra, and the contrast between these heterogeneities and the background photosphere is strong in the ultraviolet regime \citep{oshagh2014impact,llama2015transiting}.  To consider how our results for Titan might translate to other star-planet systems, we note that the transit features we observe span 6--12 scale heights.  For a sub-Neptune with a gravity of 10~m~s$^{-2}$ and an atmospheric temperature of 500~K, this range of scale heights translates to 100--200~km for a water-dominated atmosphere and 1,000--2,000~km for a H$_2$/He-dominated atmosphere.  This range of thicknesses (100--2000~km) corresponds to transit depths of 10--100~ppm for a Sun-like host and 200--3,000~ppm for a mid-M dwarf (at $0.2R_{\odot}$).  For comparison, \citet{sing2019hubble} reach uncertainties of $\sim\!20$ parts per thousand in 1~nm bins for WASP-121b with {\it HST}/STIS  while \citet{wakeford2020uv} obtain 100--1,000~ppm uncertainties in 10~nm bins for HAT-P-41b with WFC3/UVIS aboard {\it HST} --- both at longer ultraviolet wavelengths than studied here.  Thus, while the quality of atmospheric characterization obtained from {\it Cassini}/UVIS-derived transit spectra of Titan may be out of reach for {\it HST}, it may be that next-generation ultraviolet-capable space telescopes (e.g., the Large UltraViolet-Optical-InfraRed [LUVOIR; \citet{robergeetal2018}] surveyor, or the Habitable Exoplanet Observatory [HabEx; \citet{gaudietal2018}]) could better leverage exoplanet transit observations at ultraviolet wavelengths.

%
\section{Conclusions}
%

In this paper we have investigated the interplay between absorption by atmospheric hydrocarbon and organo-nitrile species and haze extinction in an occultation-derived ultraviolet transit spectrum of Titan.  Critically, Titan's complex atmospheric chemical structure may be analogous to hazy exoplanet atmospheres rich in hydrocarbon species.  Our findings are summarized as follows:

\begin{itemize}

    \item Despite extensive haze structures in Titan's atmosphere, numerous ultraviolet molecular gas absorption features are easily detected in our occultation-derived transit spectrum.
    
    \item Application of various haze parameterizations within our atmospheric retrieval framework to our ultraviolet transit spectrum of Titan all resulted in inferred gas column densities that are consistent with previous analyses of {\it Cassini}/UVIS observations.

    \item Comparison of our clearsky and hazy retrievals using the Bayesian Information Criterion strongly supports the exclusion of additional haze parameterizations.  Our model comparisons suggest that the addition of the haze parameterizations to exoplanets should be done with care, as unnecessary haze treatments might lead to poorer constraints on atmospheric parameters and increased computation time.

    \item Our transit analyses may provide weak evidence for the detection in transit of a known detached haze layer near 500 km above Titan's solid body radius. Future work addressing wavelength dependent haze extinction and/or a more sophisticated treatment of haze vertical distribution may provide more concrete transit spectral evidence for detached haze layers.
    
\end{itemize}

\acknowledgements{TDR gratefully acknowledges support from NASA's Exoplanets Research Program (No.~80NSSC18K0349) and Exobiology Program (No.~80NSSC19K0473), as well as the Nexus for Exoplanet System Science and NASA Astrobiology Institute Virtual Planetary Laboratory (No.~80NSSC18K0829).}

%
\bibliography{refs}{}
\bibliographystyle{aasjournal}
%
\pagebreak

\begin{table}[h!]
\vspace{2.5cm}
\begin{rotatetable}

    \centering
    \def\arraystretch{1.5}
    \begin{tabular}{|c c c c c c|}
    \hline
    Species & Clear & One-Parameter Haze & Two-Parameter Haze & \citet{koskinen2011mesosphere} & \citet{shemansky2005cassini} \\
    \hline
    \hline
    CH$_{4}$ & $(8.6 \pm 0.9) \times 10^{17}$ & $(8.6 \pm 0.9) \times 10^{17}$ & $(8.9 \pm 0.9) \times 10^{17}$ & $(1.60 \pm 0.03) \times 10^{18}$ & $(2.5 \pm 0.5) \times 10^{18}$ \\
    C$_{2}$H$_{2}$ & $(3.4 \pm 0.2) \times 10^{16}$ & $(3.4 \pm 0.2) \times 10^{16}$ & $(3.4 \pm 0.2) \times 10^{16}$ & $(3.40 \pm 0.05) \times 10^{16}$ & $(6 \pm 2) \times 10^{16}$ \\
    C$_{2}$H$_{4}$ & $(2.54 \pm 0.06) \times 10^{16}$ & $(2.53 \pm 0.07) \times 10^{16}$ & $(2.56 \pm 0.06) \times 10^{16}$ & $(2.50 \pm 0.02) \times 10^{16}$ & $(3.2 \pm 0.7) \times 10^{16}$ \\
    C$_{2}$H$_{6}$ & $(1.02 \pm 0.08) \times 10^{17}$ & $(1.02 \pm 0.08) \times 10^{17}$ & $(1.02 \pm 0.08) \times 10^{17}$ & $< 2.7 \times 10^{16}$ & $(4 \pm 2) \times 10^{16}$ \\
    C$_{4}$H$_{2}$ & $(3.5 \pm 0.3) \times 10^{15}$ & $(3.5 \pm 0.3) \times 10^{15}$ & $(3.6 \pm 0.3) \times 10^{15}$ & $(1.70 \pm 0.05) \times 10^{15}$ & $(2.5 \pm 0.9) \times 10^{15}$ \\
    C$_{6}$H$_{6}$ & $(3.1 \pm 0.7) \times 10^{14}$ & $(3.1 \pm 0.7) \times 10^{14}$ & $(3 \pm 1) \times 10^{14}$ & $(7.8 \pm 0.3) \times 10^{14}$ & - \\
    HCN & $(5.2 \pm 0.8) \times 10^{16}$ & $(5.1 \pm 0.9) \times 10^{16}$ & $(5.2 \pm 0.9) \times 10^{16}$ & $(2.3 \pm 0.2) \times 10^{16}$ & $(5 \pm 2) \times 10^{16}$\\
    HC$_{3}$N & $(3.3 \pm 0.2) \times 10^{15}$ & $(3.3 \pm 0.3) \times 10^{15}$ & $(3.3 \pm 0.3) \times 10^{15}$ & $(2.40 \pm 0.06) \times 10^{15}$ & - \\
    \hline
    \end{tabular}
    
    \caption{Calculated slant column densities (cm$^{-2}$) at an altitude of 700 km above Titan's solid body radius compared to previous UVIS studies.}
    
    \label{tab:comparison}

\end{rotatetable}

\end{table}

\end{document}